# Near-field Surface Waves in Few-Layer MoS$_2$

Viktoriia E. Babicheva[1], Sampath Gamage[2], Li Zhen[3], Stephen B. Cronin[3], Vladislav S. Yakovlev[4], and Yohannes Abate[2*]

[1]College of Optical Sciences, University of Arizona, Tucson, AZ 85721, USA

[2]Department of Physics and Astronomy, University of Georgia, Athens, Georgia 30602, USA

[3]Viterbi School of Engineering, University of Southern California, Los Angeles, CA 90089, USA

[4]Max Planck Institute of Quantum Optics, Hans-Kopfermann-Straße 1, Garching 85748, Germany

**Abstract**. Recently emerged layered transition metal dichalcogenides have attracted great interest due to their intriguing fundamental physical properties and potential applications in optoelectronics. Using scattering-type scanning near-field optical microscope (s-SNOM) and theoretical modeling, we study propagating surface waves in the visible spectral range that are excited at sharp edges of layered transition metal dichalcogenides (TMDC) such as molybdenum disulfide and tungsten diselenide. These surface waves form fringes in s-SNOM measurements. By measuring how the fringes change when the sample is rotated with respect to the incident beam, we obtain evidence that exfoliated MoS$_2$ on a silicon substrate supports two types of Zenneck surface waves that are predicted to exist in materials with large real and imaginary parts of the permittivity. In addition to conventional Zenneck surface waves guided along one interface, we introduce another Zenneck-type mode that exists in multilayer structures with large dissipation. We have compared MoS$_2$ interference fringes with those formed on a layered insulator such as hexagonal boron nitride where the small permittivity supports only leaky modes. The interpretation of our experimental data is supported by theoretical analysis. Our results could pave the way to the investigation of surface waves on TMDCs and other van der Waals materials and their novel photonics applications.

Layered materials, such as black phosphorous, transition metal dichalcogenides (TMDCs), and hexagonal boron nitride (hBN) are promising for a wide range of applications in optoelectronics. The optical properties of these materials at the nanoscale have attracted a great interest in recent years.[1-10] Surface waves that localize and guide energy along a layer of such a material can be used for efficient light manipulation on the nanoscale.[11] In the visible spectral range, TMDCs possess large real and imaginary parts of the permittivity, which is a prerequisite for the existence of Zenneck waves. These waves were predicted by Zenneck and Sommerfeld to exist at interfaces where both media possess positive real parts of the dielectric permittivity (Re[$\epsilon$]>0), but at least one of the media is lossy (Im[$\epsilon$]>0).[12-13] In the beginning of the last century, the propagation of radio waves along the surface of water was the main motivation for research on Zenneck waves.[14-16] While the existence of Zenneck waves follows from Maxwell's equations, contradicting claims on the observation of Zenneck waves in certain materials were reported.[11, 17-24] Experimentally, demonstration of Zenneck waves is challenging since any practical source of electromagnetic waves located at an interface between two media produces a mixed field composed of surface and bulk waves. Although direct imaging of Zenneck waves is still missing, there are a few experimental works on the observation of Zenneck waves.[22, 25-26] In recent years, the growing number of emergent layered materials, such as TMDCs, offer an excellent hunting ground for Zenneck waves in the optical spectral range.

Here, we report on the excitation and observation of Zenneck waves in the optical spectral range at the interfaces of exfoliated TMDC ($MoS_2$ or $WSe_2$) and a silicon substrate. Both TMDC and silicon are dissipative, and the Si-TMDC-air structure generally supports several guided modes. We are particularly interested in two *p*-polarized modes that have relatively low optical losses and propagate over several micrometers: Zenneck surface wave, mainly localized at TMDC/air interface, and the introduced here Zenneck-type mode, which is mainly localized at the interface between TMCD layer and the substrate. We use a scattering-type scanning near-field optical microscope (s-SNOM), which provides not only high-resolution images of nanostructures [27-39] but also serves as a tool to image all possible surface waves and waveguide modes.[8, 28, 40] s-SNOM records fringes due to the superposition of incident, reflected, scattered, and guided light at a sample surface. One of the possible mechanisms of fringe formation routinely used in s-SNOM is the interference patterns of surface polaritons (plasmon or phonon) launched by the tip and reflected at sample boundaries.[4, 28, 40-41] Sample edges also serve as efficient scatterers for both incident light and surface waves. We distinguish between different mechanisms of fringe formation by analyzing how the fringe pattern depends on the orientation of the sample with respect to the incident light. Interference fringes formed by tip-launched waves returning to the tip do not depend on the orientation, while fringes formed by waves launched at sample boundaries do. We identify contributions from various mechanisms of fringe formation by applying a two-dimensional (2D) Fourier analysis to SNOM images. We can explain our measurements as interference between light that experienced a single scattering event at the tip and light that reached the detector after being scattered by the tip and a sample edge. We do not, however, observe fringes related to tip-launched-tip-scattered waves such as those reported in the mid-infrared spectral region on graphene or hBN.[4, 28, 40-41]

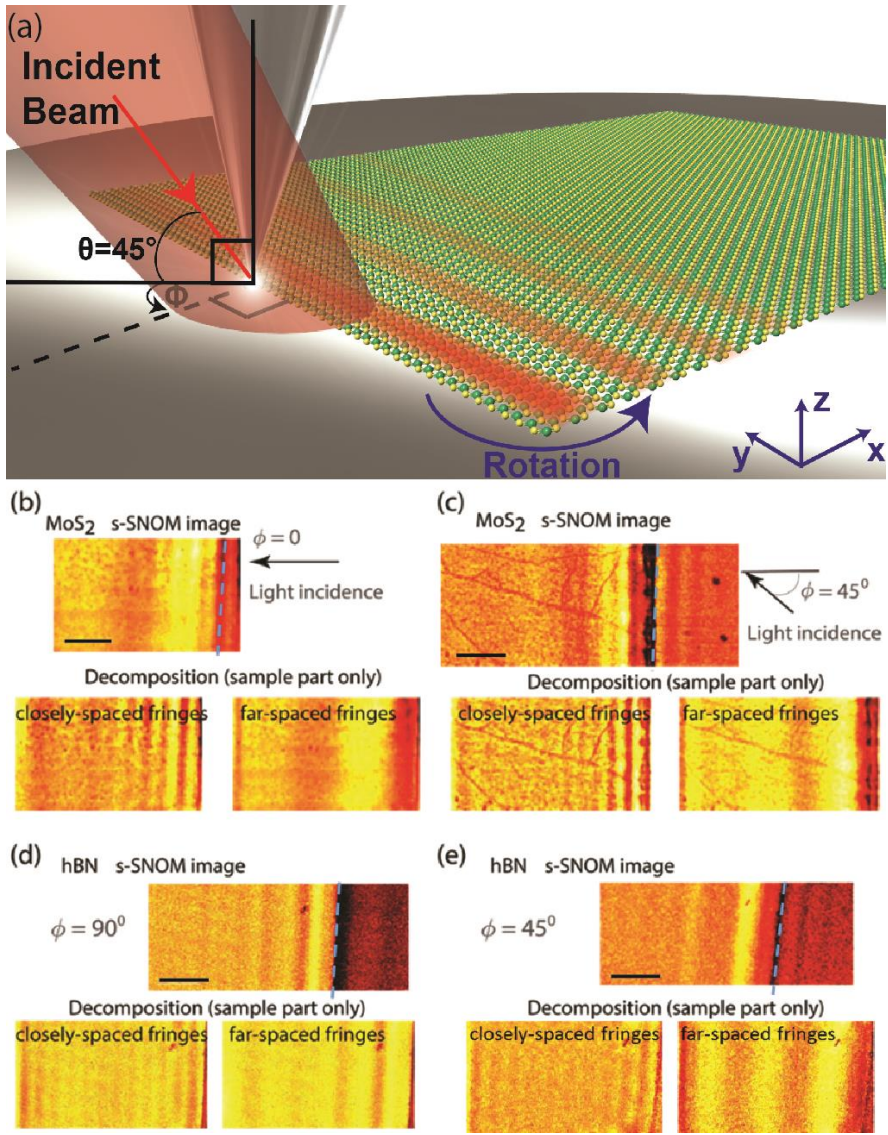

**Fig. 1.** Nano-imaging of surface waves in MoS$_2$ and hBN. (a) Artist's rendition of the experimental configuration. The blue curved arrow depicts the counterclockwise sample rotation. The tip axis and the laser illumination direction are indicated by straight arrows; these directions were fixed throughout our experiments. Sample rotation angle $\phi$ is measured on the x-y plane (sample plane) between the projection of the laser-beam axis onto the plane and the normal to the sample edge. (b)-(e) Real-space s-SNOM images of MoS$_2$ and hBN flakes: (b) MoS$_2$ flake image taken at rotation angle $\phi = 0^0$; (c) MoS$_2$ flake at $\phi = 45^0$; (d) hBN flake at $\phi = 90^0$; (e) hBN flake at $\phi = 45^0$. Fringes in an s-SNOM image can be decomposed into closely- and far-spaced components by applying spatial Fourier filtering. Fringe spacing depends on the sample rotation angle. The scans have a size of 4 x 2.5 μm$^2$. They were made for wavelength $\lambda_0 = 632$ nm, MoS$_2$ sample thickness ~130 nm, and hBN thickness ~150 nm. Scale bars in b-e represent 1 μm.

**Results and Discussion**

Figure 1a shows a schematic diagram of the s-SNOM experimental configuration. A sharp metallic cantilevered probe tip is illuminated by a laser at an angle of $\theta = 45^0$, which is measured between the beam axis and a normal to the sample surface. The sample edge surrounds an exfoliated flake (e.g., MoS$_2$) on the Si substrate. Sample rotation angle $\phi$ is measured from the projection of the incident beam axis onto the x-y plane (sample plane) and in-plane normal to the sample edge; increasing $\phi$ corresponds to the counterclockwise rotation of the sample as shown by the blue curved arrow. The scattered field from the tip-sample interface is collected using phase modulation (pseudoheterodyne) interferometry and demodulation of the detector signal at higher harmonics of the tip resonance frequency. In Fig. 1b-e, we show 4$^{th}$ harmonic amplitude images of two representative exfoliated semiconductor MoS$_2$ and insulator hBN on Si substrate. The excitation HeNe laser (wavelength, $\lambda_0$ = 632 nm) is p-polarized. Images taken at several sample rotation angles $\phi= 0^0, 45^0$, and $90^0$ are shown in Fig. 1b-e (images at additional rotations angels are also shown in the Supporting Information, Fig. S1). In these near-field images, interference fringes are visible on both samples (hBN and MoS$_2$); they are also visible on the Si substrates. These facts suggest that at least one of the mechanisms of fringe formation is material-independent. Experiments on Au films also show (Supporting Information, Fig. S2) similar results. In all of these materials, the spacing of interference fringes is strongly dependent on the sample rotation angle $\phi$ as indicated in Fig. 1. By applying low- and high-pass spatial Fourier filters, we decompose s-SNOM images into closely- and far-space fringes, see Fig. 1b-e. These two components are simultaneously present in each unfiltered image.

To determine the origin of the interference fringes and identify various possible waves, we analyzed their dependence on the sample rotation angle. Figures 2a,b show the dependence of the inverse fringe spacing, $2\pi/\Delta x$, on the sample rotation angle, $\phi$, for different samples supporting various kinds of surface waves. To analyze these dependencies, let us consider an incident electromagnetic wave of wavelength $\lambda_0$. When the wave is scattered by a sample edge, waves propagating along the sample surface emerge; let $\lambda_s$ be the wavelength of such a wave. For a Zenneck wave that is bound to the surface of the sample, $\lambda_s > \lambda_0$, which is one of the basic properties of Zenneck waves (guided waves usually have effective wavelength smaller than those of free-propagating waves; Zenneck modes are an exception allowed by strong absorption in one of the media). For the wave that is not bound to the surface (cylindrical free-space wave), $\lambda_s = \lambda_0$; for a wave guided within the TMDC layer (see the description of Zenneck-type mode below), $\lambda_s < \lambda_0$. When a wave propagating along the surface scatters at the s-SNOM tip and reaches the detector, it interferes with the incident light scattered only by the tip, without the edge being involved. The following formula estimates the fringe spacing in the case where a laser beam simultaneously illuminates the s-SNOM tip and the sample edge:

$$\Delta x_1 = \frac{\lambda_0}{\left|\sin\theta\cos\phi - \sqrt{(\lambda_0/\lambda_s)^2 - (\sin\theta\sin\phi)^2}\right|}. \tag{1}$$

We refer to section Interference Model below for details on the model and the derivation of Eq. (1). For the cylindrical wave, which exists even if the sample does not support any guided or surface waves, $\lambda_s = \lambda_0$ simplifies Eq. (1) to

$$\Delta x_2 = \frac{\lambda_0}{\left|\sin\theta\cos\phi - \sqrt{1-(\sin\theta\sin\phi)^2}\right|}. \tag{2}$$

If there is a surface wave and a cylindrical wave, their interference with each other results in fringes spaced by

$$\Delta x_3 = \frac{\lambda_0}{\left|\sqrt{(\lambda_0/\lambda_s)^2 - (\sin\theta\sin\phi)^2} - \sqrt{1-(\sin\theta\sin\phi)^2}\right|}. \tag{3}$$

The three types of interference fringes are shown in Figs. 2a and 2b with black, green, and red curves, respectively, where we superimpose $2\pi/\Delta x$ evaluated with Eqs. (1-3) on the experimental data.

We observed the dependence of interference fringes on the sample rotation angle, $\phi$, for both TMDC and hBN samples. We found the main contribution comes from the interference of the incident light with the cylindrical wave. The propagation of these edge-scattered cylindrical waves in air is practically independent of the sample's optical properties. In the study of surface waves, this contribution can be considered as an artifact. Such cylindrical waves attracted a significant interest in earlier studies of propagating surface plasmons and extraordinary optical transmission associated with them.[42-44] In the present work, we do not study these cylindrical waves in detail; instead, we point out their prominent and often misleading effect on experimental near-field interference image formation and interpretation not only in the visible but also in the infrared frequency range. For example, we have performed imaging on TMDC samples in the mid-infrared wavelength range, $\lambda$ = 4-10 µm), (other excitations, such as interband or excitonic transitions are not expected in this energy range), fringes are only caused by the interference of the incident beam with edge-scattered cylindrical waves (see WSe$_2$ scans in Fig. S3 of the Supporting Information). These infrared images show that the artifact fringes can extend from the sample edge on over several tens of micrometers (Supporting Information, Fig. S4). Also, in our experiments, both TMDCs and hBN are deposited on top of silicon rather than on standard SiO$_2$/Si substrates. A SiO$_2$ layer with a thickness of several hundred nanometers would support waveguide modes confined between the TMDC and the underlying Si, and these long-range guided waves may prevent the observation of the surface waves considered here.[8]

To distinguish between surface and waveguide modes, we numerically solved the dispersion equation for the air/MoS$_2$/silicon three-layer structure. We found solutions that correspond to Zenneck surface waves (Fig. 2c,g), which become conventional Zenneck waves in the limit of an infinitely thick TMDC layer (Fig. 2e), as well as Zenneck-type modes (Fig. 2d,h), where both interfaces participate in the formation of a guided wave and mode tail in silicon is similar to that one for MoS$_2$/silicon (Fig. 2f). Both types of modes are enabled by the large dissipation in MoS$_2$ and Si. These multilayer structures also support other two types of the modes: waveguide modes and leaky waves.[45-47] Waveguide modes are confined in the TDMC layer because the real part of its permittivity is larger than those of the Si substrate and the air. These modes have a field maximum inside the MoS$_2$ layer, and their wavelength is < 170 nm. Due to high nonradiative losses, the attenuation lengths for waveguide modes are ≤ 0.25 µm, which makes them irrelevant to our measurements. In contrast to Zenneck and waveguide waves, which are localized either at an interface or in the high-index layer, the leaky waves are solutions of dispersion equations where the electric field exponentially increases away from one of the boundaries. Not being bound to the TMDC layer, these modes possess very high radiative losses and decay much faster than Zenneck and waveguide modes. Therefore, we do not consider them in the further TMDC analysis.

Note also that Zenneck- and Zenneck-type waves are different from *s*-polarized exciton-polariton modes in TMDC monolayer in symmetric cladding[48], which are also layer-bound waves. In addition to the polarization, these two kinds of modes differ in their wavelength: while the effective wavelength of a Zenneck mode at the boundary between two media exceeds the wavelengths of plane waves in each of these media, the wavelength of an exciton-polariton mode propagating along a TMDC monolayer is smaller than that of an electromagnetic wave freely propagating in the ambient medium. Also, nearly-equal refractive indices of materials at both sides of a monolayer are essential to exciton-polariton modes[48].

The difference between Zenneck and waveguide modes is evident from the guiding role of either one or both TMDC interfaces. The in-plane component of the electric field of the Zenneck surface wave is maximal at the interface between a lossy medium, $MoS_2$, and non-lossy one, air (Fig. 2c). The wave significantly extends into the non-lossy medium, and its wavelength is larger than the free-space wavelength: $\lambda_s = \lambda_{ZW} = 642$ nm $> \lambda_0 = 632$ nm. At this excitation wavelength, the $MoS_2$/silicon interface also supports Zenneck waves (Fig. 2f) because $MoS_2$ is much lossier than silicon; a Zenneck wave also exists at the air/silicon interface [22-23], that is, at the bare substrate.

Since the wavelength of the Zenneck surface wave exceeds $\lambda_0$ by as little as 2%, our data does not allow us to clearly resolve Zenneck surface waves from cylindrical waves, for which $\lambda = \lambda_0$. We do, however, resolve the Zenneck-type wave. Its profile is shown in Fig. 2d,h for a 130-nm-thick $MoS_2$ flake exfoliated on a Si substrate, where the mode has a wavelength of 260 nm. We refer to this mode as a Zenneck-type wave because it is similar to the Zenneck wave that exists at a single interface between silicon and $MoS_2$ (Fig. 2f). The profile of the Zenneck-type mode extends predominantly into the silicon substrate, and the field maximum is located at the silicon/$MoS_2$ interface (Fig. 2d). Mode properties are summarized in Table 1. According to our calculations, the Zenneck and Zenneck-type modes have propagation lengths of 38 and 2.7 μm, respectively (see Table 1), which is in striking contrast to a waveguide mode with the propagation length below 0.25 μm. Waveguide mode losses are strictly proportional to the material losses of TMDC even for the imaginary part of permittivity comparable to real part, while there is no such relationship for Zenneck surface waves. Mode propagation contants are used in Eq. (1)-(3) to plot interference results in Fig. 2a,b and show good agreement with experimental measurements.

Table 1. Comparison of propagation constant $k^{(s)}$, effective wavelength $\lambda_s$, mode propagation length $L_s$, and depth of mode penetration into the air $w_a$ and silicon $w_{Si}$ for the four considered waves.

| Wave | $k^{(s)}$, μm$^{-1}$ | $\lambda_s$, nm | $L_s$, μm | $w_a$, μm | $w_{Si}$, μm |
|---|---|---|---|---|---|
| Incident | 9.94 | 632 | NA | NA | NA |
| Cylindrical | 9.94 | 632 | NA | NA | NA |
| Zenneck | 9.85 | 642 | 38 | 5.8 | 6.4 |
| Zenneck-type mode | 24.2 | 260 | 2.7 | 0.04 | 0.57 |

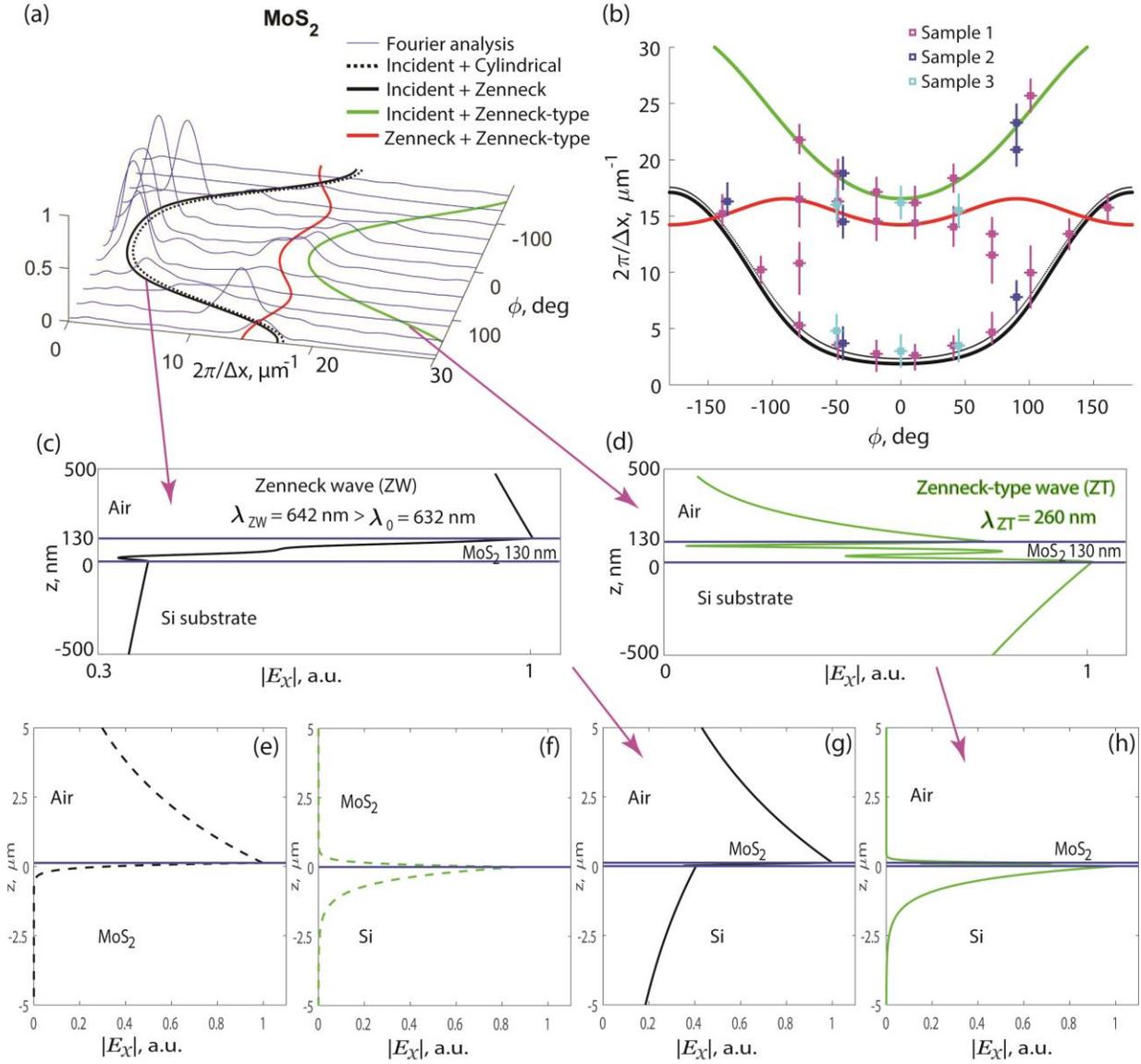

**Fig. 2.** Fringe-spacing analysis for MoS$_2$. (a) Fourier analysis of the s-SNOM images measured over the MoS$_2$ part of sample 1 for different sample rotation angles, $\phi$. The analysis results are overlaid with the inverse fringe spacing $2\pi/\Delta x$ calculated with Eqs. (1)-(3) for the interfering light waves: the incident light, cylindrical free-space waves, Zenneck surface waves, and Zenneck-type modes of highly dissipative MoS$_2$. (b) The dots and the error bars represent the positions of the main peaks in the Fourier analysis and their half-maximum widths, respectively. Here, we show data for three samples. The black, green, and red theoretical curves are the same as those in (a). (c)-(h) Mode profiles of Zenneck surface and Zenneck-type waves, as well as modes at a single interface: the plots show $|E_x|$, which is the absolute value of the $E$-field component parallel to the surface. In (c), the Zenneck surface wave has a wavelength of $\lambda_{ZW} = 642$ nm, which is larger than the free-space wavelength of the incident light $\lambda_0 = 632$ nm. The mode is largely extended into the surrounding air. In (d), Zenneck-type mode with $\lambda_{ZT} = 260$ nm and largely extended into the Si substrate. (e) The mode profile at the MoS$_2$/air interface is similar to that of the Zenneck wave in the three-layer structure and has a comparable mode penetration depth into air. (f) The mode profile at the MoS$_2$/Si interface resembles that of the Zenneck-type mode with a similar penetration depth into silicon. (g) and (h) are the same as (c) and (d) but at a large scale along $z$-axis.

In contrast to conventional Zenneck waves, Zenneck-type modes are only supported in a spectral range where both real and imaginary parts of the TMDC permittivity have sufficiently large values, while the limiting values sensitively depend on the layer thickness. Here we show an example of this mode for a particular MoS$_2$ permittivity and thickness considered in our work. Various studies report different values for MoS$_2$ permittivity [49-53] that strongly depend on fabrication conditions. The wavelength of our study ($\lambda_0 = 632$ nm) overlaps with the exciton resonance energy in MoS$_2$ (1.8 - 2 eV), which affects the material permittivity. In our model, we estimated the MoS$_2$ permittivity as ε ~ 30 + 8.5i (Ref. [51]) to reproduce the measured fringe spacing.

With the same approach and technique employed for fringes on MoS$_2$, we study hBN samples, showing the results in Fig. 3. We used mechanically exfoliated hBN on a Si substrate as a control sample to demonstrate interference patterns in the absence of Zenneck waves. The band gap of hBN is at about 5.2 eV (wavelength 238 nm), at a wavelength of $\lambda_0 = 632$ nm; hBN is a dielectric with a moderate permittivity: the real part is about 4 - 5 and imaginary is about 0.5.[54-55] Our sample is ~150 nm thick and, because of the small optical thickness, does not support any non-leaky modes bound to the material layer. Neither isotropic nor anisotropic permittivity in the range mentioned above supports bound surface modes, so only leaky waves can be excited. Figure 3 shows the results of calculations for $\varepsilon_{hBN} = 4 + 0.5i$. Therefore, for the hBN sample, we observe interference fringes that correspond to two waves only: cylindrical and leaky. The leaky wave has a wavelength of $\lambda_{LW} = 422$ nm, but the mode is not localized at the hBN sample and has a field component that increases away from the hBN/silicon interface (see inset in Fig. 3). Despite high radiative losses, this mode contributes to interference fringes within several micrometers from the hBN edge. The absence of guided Zenneck and Zenneck-type modes makes the leaky mode visible in the interference pattern on the hBN layer.

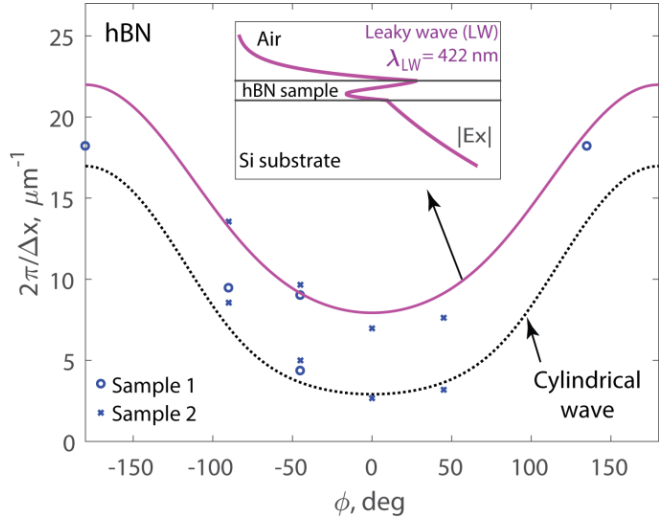

**Fig. 3.** Fringe spacing analysis for hBN: Inverse fringe spacing $2\pi/\Delta x$ dependence on the sample rotation angle $\phi$. Lines correspond to the model Eqs. (1)-(2) of interference between incident light and either cylindrical or leaky waves excited at the edge of hBN. Inset: The leaky wave mode profile shows $|E_x|$, which is the absolute value of the $E$-field component parallel to the surface; $|E_x|$ increases away from the hBN/silicon interface.

**Conclusion**

In summary, we have studied surface waves in a representative TMDC layer (i.e., $MoS_2$) mechanically exfoliated on a bare Si substrate at optical frequencies using s-SNOM and a theoretical model at various sample-rotation angles. We have demonstrated the existence of Zenneck-type waves that are enabled by the high dissipation of $MoS_2$, and we have imaged their interference in real space. We note that the Zenneck-type mode that we analyzed here disappears if the thickness of the $MoS_2$ layer is increased; however, this is not a general property of Zenneck-type modes. Proximity to a cutoff increases the propagation length of Zenneck-type modes. As a control experiment, we have used mechanically exfoliated hBN (representative insulator) on a Si substrate to demonstrate interference patterns due to hBN edge-reflected cylindrical waves that propagate in the air for tens of micrometers without regard to sample optical properties. Propagating surface electromagnetic waves are at the heart of exciting research and novel applications in nanooptics. We anticipate that the presence of Zenneck-type surface waves that we have studied in modern quantum materials such as layered TMDC will offer yet new opportunities in nanophotonics. Since Zenneck-type modes are sensitive to the properties of the interface, they can be used to study surface and interface physics and chemistry.

**Methods**

The microscope is a commercial s-SNOM system (neaspec.com). A probing linearly p-polarized HeNe laser (632 nm) or QCL laser (mid IR) illuminates the tip and the sample edge at an angle of $45^0$ to the sample surface. The scattered field is acquired using a phase modulation (pseudoheterodyne) interferometry. The background signal is suppressed by vertical tip oscillations at the mechanical resonance frequency of the cantilever ($f_0 \sim 285$ kHz) and demodulation of the detector signal at higher harmonics $nf_0$, $n = 2, 3, 4$, of the tip resonance frequency. The combined scattered field from the tip and the reference beam pass through a linear polarizer, which further selects the p polarization of the measured signal for analysis. The TMDC and hBN samples were prepared using mechanical exfoliation on pre-cleaned Si wafer and imaged using s-SNOM at various sample-rotation angles $\phi$ at a wavelength of $\lambda_0 = 632$ nm and in the mid-IR range. Typical flake sizes are about 20-40 μm and the imaged areas were about 5 x 5 μm².

**Interference Model**

Consider an incident wave with the real part of its wavevector

$$\text{Re}[\boldsymbol{k}^{(0)}] = \left(k_x^{(0)}, k_y^{(0)}, k_z^{(0)}\right). \tag{M1}$$

($\boldsymbol{k}^{(0)}$ is real unless the incidence medium absorbs light). This wave excites a surface wave propagating along the sample in the direction away from the sample edge (sample boundaries as illustrated in Fig. 1). Let the wavevector of the surface wave be $\text{Re}[\boldsymbol{k}^{(s)}] = \left(k_x^{(s)}, k_y^{(s)}, k_z^{(s)}\right)$.

Along the edge, the phases of the two waves must coincide. Therefore,

$$k_y^{(0)} = k_y^{(s)}. \tag{M2}$$

We have verified that the imaginary part of $\boldsymbol{k}^{(s)}$ has a negligible effect on the fringe spacing for our parameters. Under this premise, the spacing between the fringes formed by the interference of the incident light and the surface wave equals

$$\Delta x = \frac{2\pi}{\left|k_x^{(0)} - k_x^{(s)}\right|}. \tag{M3}$$

Note that the fringes are parallel to the edge even if $\phi \neq 0$. For materials isotropic in the $xy$ plane, the $x$-component of $\text{Re}[\boldsymbol{k}^{(s)}]$ can be found from $\sqrt{\left(k_x^{(s)}\right)^2 + \left(k_y^{(s)}\right)^2} = 2\pi/\lambda_s$, where $\lambda_s$ is the wavelength of the surface mode measured along the surface (even though, in general, $k_z^{(s)} \neq 0$, this component of the wavevector plays no role at $z = 0$). So,

$$k_x^{(s)} = \sqrt{\left(\frac{2\pi}{\lambda_s}\right)^2 - \left(k_y^{(s)}\right)^2} \tag{M4}$$

We expect similar results for equal positive and negative angles $\phi$, yet we have performed a rotation in both directions and kept this information of the measured data on the plot.

Substituting Eq. (M4) into (M3), as well as using $k_x^{(0)} = 2\pi \sin(\theta)\cos(\phi)/\lambda_0$ together with $k_y^{(0)} = 2\pi \sin(\theta)\sin(\phi)/\lambda_0$, we obtain Eq. (1).

If there are two types of waves propagating along the surface, their interference produces fringes spaced by

$$\Delta x = 2\pi / \left| k_x^{(s1)} - k_x^{(s2)} \right|. \qquad (M5)$$

Note that fringe spacing would be different if the laser beam were tightly focused on the tip, without illuminating the sample edge.[8]

## Acknowledgments

This work is supported by the National Science Foundation CAREER grant no. 1553251 (Y.A. and S.G.). V.B. acknowledges support by the Air Force Office of Scientific Research grant number FA9559-16-1-0172. This research was supported in part by the Department of Energy Award No. DE-FG02-07ER46376 (Z.L.). The authors are grateful to Marquez Howard for assistance in the preparation of figures and Mark Stockman for initial useful discussions.

## Author information Notes

Corresponding Author:     Dr. Yohannes Abate

Email: yabate@physast.uga.edu

**Supporting Information:** Near-field images of surface waves on $MoS_2$ for additional sample rotation angles, near-field images of surface waves on $WSe_2$ in mid-IR range, and near-field images of surface waves on gold.

## Competing financial interests

The authors declare no competing financial interests.

For Table of Contents Use Only

**Near-field Surface Waves in Few-Layer MoS$_2$**

Viktoriia E. Babicheva, Sampath Gamage, Zhen Li, Stephen B. Cronin, Vladislav S. Yakovlev, and Yohannes Abate

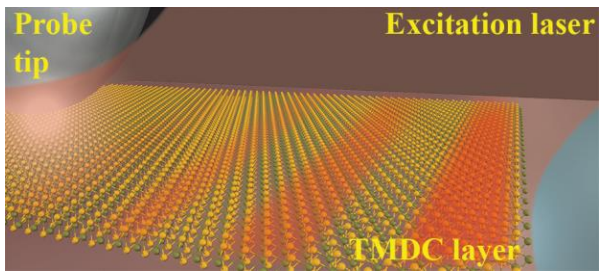

Near-field nanoimaging of propagating surface waves

# Near-field Surface Waves in Few-Layer MoS$_2$

Viktoriia E. Babicheva[1], Sampath Gamage[2], Zhen Li[3], Stephen B. Cronin[3], Vladislav S. Yakovlev[4], and Yohannes Abate[2]

[1]College of Optical Sciences, University of Arizona, Tucson, AZ 85721, USA

[2]Department of Physics and Astronomy, University of Georgia, Athens, Georgia 30602, USA

[3]Viterbi School of Engineering, University of Southern California, Los Angeles, CA 90089, USA

[4]Max Planck Institute of Quantum Optics, Hans-Kopfermann-Straße 1, Garching 85748, Germany

Number of pages: 3

Number of figures: 4

Number of tables: 0

Supporting Information. Some of the additional experimental results obtained for different incidence angles of laser on MoS2, Au and WSe2 samples at λ=632 nm and at mid-infrared wavelength range.

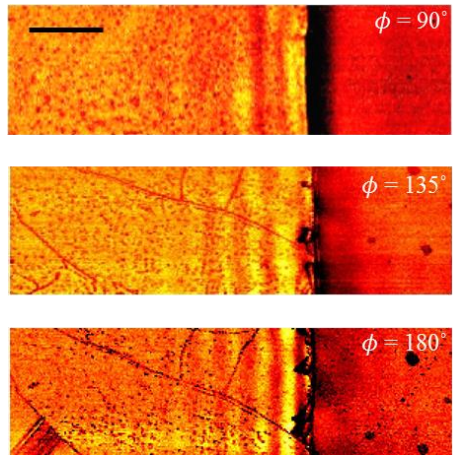

Figure S1. Real-space s-SNOM amplitude images of MoS$_2$ on Si at λ=632 nm, measured at three different light incident angles, $\phi$ = 90˚, 135˚ and 180˚. Scale bar 1 μm.

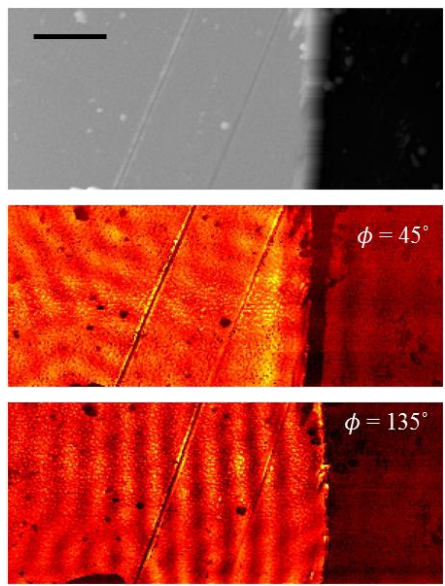

Figure S2. Topography and real-space s-SNOM amplitude images of 95 nm thick Au layer on Si at λ=632 nm, measured at two different light incident angles, $\phi$ = 45˚ and 135˚. Scale bar 1 μm.

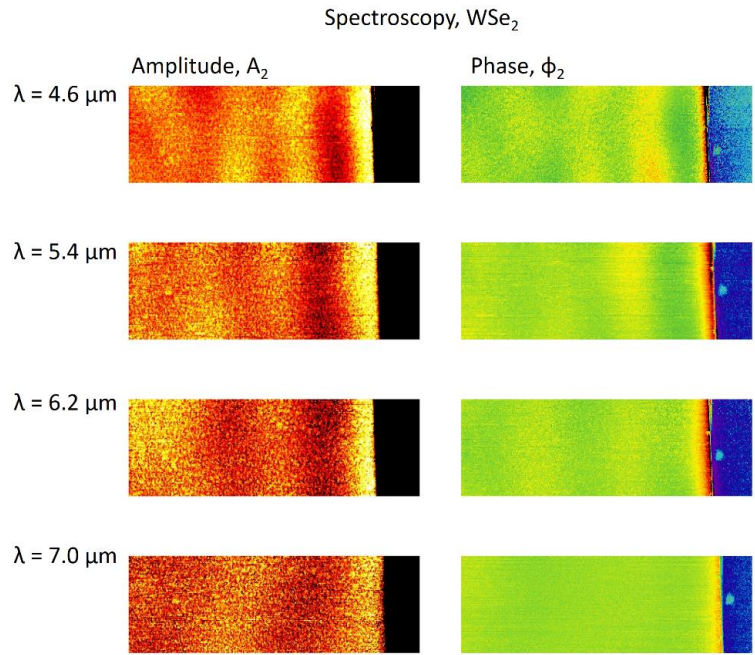

Figure S3. Real-space s-SNOM images of interference fringes at WSe$_2$ flake in mid-infrared range. They arise because of interference of incident light with the cylindrical wave scattered by the sample edge.

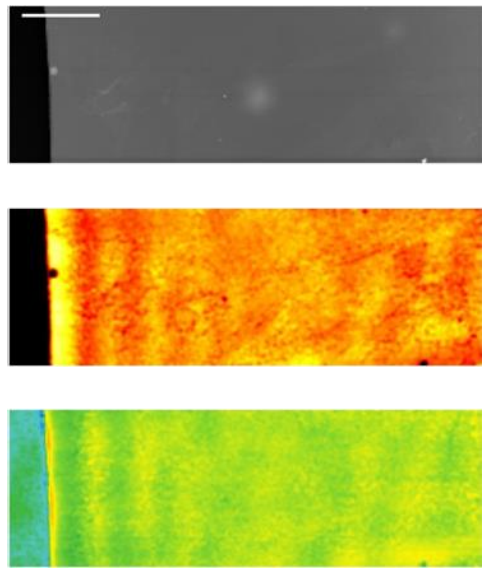

Figure S4. Topography, second harmonic amplitude and phase of a 130 nm thick WSe$_2$ flake on SiO$_2$, imaged at 4.6 μm. Scan size 30 μm x 10 μm. Scale bar 5 μm.